\documentclass[fleqn]{article}
\usepackage{./styles/iopconf1}
\usepackage[german,english]{babel}
\usepackage{rotating}
\usepackage{epsfig}
\usepackage[centertags]{amsmath}
\usepackage{amssymb,amscd}
\usepackage{./styles/amsatdef}
\usepackage{./styles/extrafonts}
\usepackage{tabularx}
\usepackage{multirow}
\usepackage{./styles/axodraw}
\usepackage{dcolumn}
\usepackage{xspace}
\usepackage{calc}
\usepackage{./styles/mcite}
\usepackage{cite}
\def\zero{{\scriptscriptstyle 0}}

\def\none{---}

\def\Z0{{Z^\zero}}

\def\Tev{{\text{Te}@!@!@!@!\text{V\/}}}

\def\gev{{\,\text{Ge}@!@!@!@!\text{V\/}}}

\def\tev{{\,\text{Te}@!@!@!@!\text{V\/}}}
\def\pb{\,\text{pb}}

\def\pbi{\,\text{pb}^{-1}}

\def\BR{{\rm BR}}

\def\DO{{D{\O}}\xspace}

\def\LL{{\scriptscriptstyle{\rm LL}}}

\def\LQsub{{\scriptscriptstyle{\rm LQ}}}
\def\LQ{{\rm LQ}}
\def\LR{{\scriptscriptstyle{\rm LR}}}

\def\RL{{\scriptscriptstyle{\rm RL}}}

\def\RR{{\scriptscriptstyle{\rm RR}}}

\def\lim{{\rm lim}}

\def\ord#1{{\cal O}(#1)}
\def\ptmiss{{\not{\kern-0.3em P_t}}}

\def\smin{{\rm min}}

\mathchardef\qsm=63
\mathchardef\pls=43
\mathchardef\mns=512
\mathchardef\plm=518
\mathchardef\eql=61
\mathchardef\smallleft=300
\mathchardef\smallright=301
\mathchardef\perslsh=47
\mathchardef\les=316
\mathchardef\gre=318
\mathchardef\leq=532
\mathchardef\grq=533
\mathchardef\rar=545
\chardef\usc=95
\chardef\til=126

\def\sqr#1#2#3{{\vcenter{\hrule height.#3ex\hbox{\vrule width.#2ex height#1ex
    \kern#1ex\vrule width.#3ex}\hrule height.#2ex}}}

\def\angleto{\vrule width.035em height2.1ex depth-.56ex\unskip\kern-.6ex\to}
\def\perchc#1{{\raise.4ex\hbox{$\mkern4mu#1{\it\perslsh}_
             {\mkern-5mu\scriptscriptstyle{{\rm o}\!{\rm o}}}^
             {\mkern-12.8mu\scriptscriptstyle{\rm o}}$}}}

\def\widebar#1{\mkern1.5mu\overline{\mkern-1.5mu#1\mkern-1.mu}\mkern1.mu}
\catcode`\@=11 
\def\parenbar{\mathpalette\p@renb@r}
\def\p@renb@r#1#2{\vbox{%
  \ifx#1\scriptscriptstyle \dimen@.7em\dimen@ii.2em\else
  \ifx#1\scriptstyle \dimen@.8em\dimen@ii.25em\else
  \dimen@1em\dimen@ii.4em\fi\fi \offinterlineskip
  \ialign{\hfill##\hfill\cr
    \vbox{\hrule width\dimen@ii}\cr
    \noalign{\vskip-.3ex}%
    \hbox to\dimen@{$\mathchar300\hfil\mathchar301$}\cr
    \noalign{\vskip-.3ex}%
    $#1#2$\cr}}}
\catcode`\@=12 

\def\pbar{\widebar{p}}

\def\qbar{\widebar{q}}

\def\nubar{\widebar{\nu}}

\newbox\struttbox
\setbox\struttbox=\hbox{\vrule height1.65ex depth.485ex width0pt}
\def\strutt{\relax\ifmmode\copy\struttbox\else\unhcopy\struttbox\fi}
\def\stru#1#2{\relax\ifmmode\hbox{\vrule height#1 depth#2 width0pt}
\else\vrule height#1 depth#2 width0pt\fi}

\def\ronum#1{\uppercase\expandafter{\romannumeral#1}}
\def\ronuml#1{\expandafter{\romannumeral#1}}

\def\pcite#1{\protect\cite{#1}}

\def\fig#1{Fig.~\ref{fig-#1}}

\def\tab#1{Tab.~\ref{tab-#1}}

\DeclareMathAlphabet{\mathbf}{OT1}{cmr}{bx}{sl}

\catcode`\@=11 
\let\tab@penalty\relax
\newcount\tab@state
\def\bcline#1{%
  \noalign{\kern-.5\arrayrulewidth\tab@penalty}%
  \omit%
  \global\tab@state\@ne%
  \ranges\bcline@i{#1}%
  \cr%
  \noalign{\kern-.5\arrayrulewidth\tab@penalty}%
}
\def\bcline@i#1#2{%
  \ifnum#1<\tab@state\relax%
    \tab@@cr%
    \noalign{\kern-\arrayrulewidth\tab@penalty}%
    \omit%
    \global\tab@state\@ne%
  \fi%
  \@whilenum\tab@state<#1\do{%
    \hfil\tab@@tab@omit%
    \global\advance\tab@state\@ne%
  }%
  \ifnum\tab@state>\@ne%
    \kern-\arrayrulewidth%
  \fi%
  \@whilenum\tab@state<#2\do{%
    \tab@@span@omit%
    \global\advance\tab@state\@ne%
  }%
  \leaders\hrule\@height\boldarrayrulewidth\hfill%
}
\def\ranges#1#2{%
  \gdef\ranges@temp{#1}%
  \begingroup%
  \ranges@i#2 \q@delim%
}
\def\ranges@i{%
  \@ifnextchar\q@delim\ranges@done{\afterassignment\ranges@ii\count@}%
}
\def\ranges@ii{%
  \@ifnextchar-\ranges@iii{\ranges@do\count@\count@\ranges@v}%
}
\def\ranges@iii-{\afterassignment\ranges@iv\@tempcnta}
\def\ranges@iv{\ranges@do\count@\@tempcnta\ranges@v}
\def\ranges@v{%
  \@ifnextchar,%
    \ranges@vi%
    {%
      \@ifnextchar\q@delim%
        \ranges@done%
        {\tab@err@range\ranges@vi,}%
    }%
}
\def\ranges@vi,{\afterassignment\ranges@ii\count@}
\def\ranges@do#1#2{%
  \ifnum#1>#2\else%
    \expandafter\endgroup%
    \expandafter\ranges@temp%
    \expandafter{%
    \the\expandafter#1%
    \expandafter}%
    \expandafter{%
    \the#2%
    }%
    \begingroup%
  \fi%
}
\def\ranges@done\q@delim{\endgroup}
\def\ifinrange#1#2{%
  \@tempswafalse%
  \count@#1%
  \ranges\ifinrange@i{#2}%
  \if@tempswa%
    \expandafter\@firstoftwo%
  \else%
    \expandafter\@secondoftwo%
  \fi%
}
\def\ifinrange@i#1#2{%
  \ifnum\count@<#1 \else\ifnum\count@>#2 \else\@tempswatrue\fi\fi%
}
\def\tab@@cr{\cr}
\def\tab@@tab@omit{&\omit}
\def\tab@@span@omit{\span\omit}
\def\tab@checkrule#1{%
  \count@#1\relax%
  \expandafter\ifinrange%
  \expandafter\count@%
  \expandafter{\tab@xcols}%
    {\tab@checkrule@i}%
    {}%
}
\def\bhline{\noalign{\ifnum0=`}\fi\hrule \@height  
\boldarrayrulewidth \futurelet \@tempa\@xhline}
\def\@xhline{\ifx\@tempa\hline\vskip \doublerulesep\fi
      \ifnum0=`{\fi}}
\catcode`\@=12 
\newcommand{\topboldline}{\bhline\bs}
\newcommand{\midboldline}{\bs\bhline\bs}
\newcommand{\bottomboldline}{\bs\bhline}
\catcode`\@=11 
\newcounter{pict@width}
\newcounter{pict@height}
\newlength{\pict@scale}
\newcommand{\psfigadd}[4]{%
\setcounter{pict@width}{1*\ratio{#2+\pict@scale/2}{\pict@scale}}
\setcounter{pict@height}{1*\ratio{#3+\pict@scale/2}{\pict@scale}}
\setlength{\unitlength}{\pict@scale}
\hbox{\begin{picture}(\thepict@width,\thepict@height)
\put(0,0){\psfig{figure=#1,width=#2,height=#3,clip=}}
\SetScale{0.283466457}
\SetWidth{1.763889}
{#4}
\end{picture}}
}
\newcommand{\psfigror}[4]{%
\setcounter{pict@width}{1*\ratio{#2+\pict@scale/2}{\pict@scale}}
\setcounter{pict@height}{1*\ratio{#3+\pict@scale/2}{\pict@scale}}
\setlength{\unitlength}{\pict@scale}
\hbox{\begin{picture}(\thepict@width,\thepict@height)
\put(0,\thepict@height){\psfig{figure=#1,width=#3,height=#2,clip=,angle=270}}
\SetScale{0.283466457}
\SetWidth{1.763889}
{#4}
\end{picture}}
}
\newcommand{\psfigrol}[4]{%
\setcounter{pict@width}{1*\ratio{#2+\pict@scale/2}{\pict@scale}}
\setcounter{pict@height}{1*\ratio{#3+\pict@scale/2}{\pict@scale}}
\setlength{\unitlength}{\pict@scale}
\hbox{\begin{picture}(\thepict@width,\thepict@height)
\put(0,0){\psfig{figure=#1,width=#3,height=#2,clip=,angle=90}}
\SetScale{0.283466457}
\SetWidth{1.763889}
{#4}
\end{picture}}
}
\catcode`\@=12 

 {\end{list}}
 {\setlength{\topsep}{#1}
  \setlength{\itemsep}{#2}
  \begin{enumerate}}
 {\end{enumerate}}

\catcode`\@=11 
\newlength{\@fninsert}
\newlength{\@fnwidth}
\renewcommand{\@makefntext}[1]%
  {\noindent\makebox[\@fninsert][r]{\@makefnmark}\hfil%
  \parbox[t]{\@fnwidth}{\rm\noindent{#1}}}
\catcode`\@=12 
\newlength{\localtextwidth}
\catcode`\@=11 
\newsavebox{\tmpbox}
\newlength{\@captionmargin}
\newlength{\@captionwidth}
\newlength{\@captionitemtextsep}
\newlength{\@captionraggedlimit}
\newcommand{\@captionlinebreak}{\tolerance=2000}
\renewcommand{\@makecaption}[2]%
  {\def\baselinestretch{0.95}%
   \vspace{4.pt}
   \setlength{\@captionwidth}{\localtextwidth}
   \addtolength{\@captionwidth}{-\@captionmargin}
   \sbox{\tmpbox}{{\bf #1:}{\it #2}}%
   \ifthenelse{\lengthtest{\@captionwidth > \@captionraggedlimit}}%
   {\renewcommand{\@captionlinebreak}{\tolerance=2000}}%
   {\renewcommand{\@captionlinebreak}{\raggedright}}
   \ifthenelse{\lengthtest{\wd\tmpbox > \@captionwidth}}%
   {\centerline{\parbox[t]{\@captionwidth}%
   {\@captionlinebreak\normalsize%
    {\bf #1:}\hspace{\@captionitemtextsep}{\it #2}}}}%
   {\centerline{{\bf #1:}\kern1.em{\it #2}}}}
\catcode`\@=12 

\usepackage{./styles/mysidecaption}
\begin{document}
\selectlanguage{english}
\rightline{BONN-HE-99-06}
\rightline{Bonn University$\,$}
\title{Search for excited fermions and\\ leptoquarks at HERA}

\author{Ulrich F. Katz\footnote{$\!\!$supported by a grant by the German Federal
                                Ministry for Education and Science, Research
                                and Technology} (ZEUS)\\
        {\small Representing the H1 and ZEUS Collaborations}}

\affil{University of Bonn, Institute of Physics,\\ 
       Nu\ss allee 12, 53115 Bonn, Germany\\
       Email: katz@physik.uni-bonn.de}

\beginabstract
Recent results on searches for new particles at the electron-proton collider
HERA are reported. Based on roughly $40\pbi$ of $e^+p$ data taken in the years
1994--1997, the H1 and ZEUS collaborations have derived new exclusion limits for
the direct production of excited fermion states and of leptoquarks in different
decay channels, including lepton-flavor violating decays. The results of
searches for contact interactions further constrain the parameter space for such
particles and their couplings in the high-mass regime, where direct production
is kinematically prohibited. Also preliminary analyses of the $e^-p$ data taken
in 1998 and 1999 do not find signals of new physics.
\endabstract
\vspace*{-12.mm}

\section{Introduction}
\label{sec-int}

Between mid--1994 and the end of 1997 the electron-proton collider HERA at DESY
has been operated with positrons ($e^+$) at an energy of $E_e=27.5\gev$ and
protons ($p$) of $E_p=820\gev$, yielding a center--of--mass energy of $\sqrt{s}=
300\gev$. In this period, ZEUS and H1 have collected $e^+p$ data samples
corresponding to integrated luminosities of $47.7\pbi$ and $37\pbi$,
respectively. In 1998 and the first half of 1999, each experiment has taken
about $15\pbi$ of $e^-p$ data at $\sqrt s=318\gev$. This paper describes the
status of searches for signals of excited fermions, leptoquarks and contact
interactions in these data samples.  Many of the results are still preliminary,
and some have only become available after this {\sc Beyond99} conference but are
included here to provide an up-to-date overview of recent experimental results.
Searches at HERA for $R$-parity violating supersymmetry and for the production
of isolated leptons with high transverse momentum ($P_t$) are covered in
\cite{lem-99-01}.

\section{Excited fermions}
\label{sec-efer}

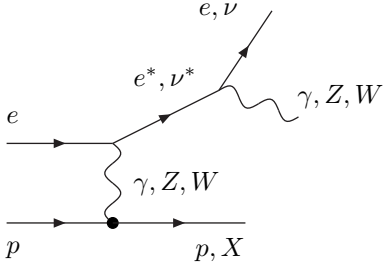
\begin{figure}[ht]
  \sidecaption
  \begin{picture}(150.,90.)(0.,0.)
  \SetOffset(0.,-20.)
  \ArrowLine(10,30)(50,30)            
  \ArrowLine(10,60)(50,60)            
  \Photon(50,30)(50,60){3}{2}         
  \GCirc(50,30){2}{0}                 
  \ArrowLine(50,30)(100,30)           
  \ArrowLine(50,60)(90,80)            
  \ArrowLine(90,80)(110,110)          
  \Photon(90,80)(120,70){3}{2}        
  \Text(10,70)[l]{$e$}                
  \Text(10,20)[l]{$p$}                
  \Text(100,20)[r]{$p,X$}             
  \Text(58,45)[l]{$\gamma,Z,W$}       
  \Text(70,80)[b]{$e^\ast,\nu^\ast$}  
  \Text(90,110)[]{$e,\nu$}            
  \Text(120,78)[l]{$\gamma,Z,W$}      
  \end{picture}
  \caption{Feynman graph of the production and subsequent decay of excited
           electrons or neutrinos in $ep$ reactions. Note that alternatively
           also an excited quark could be produced at the boson-proton vertex.
           In the case of $e^\ast$ production, elastic reactions $ep\to e^\ast 
           p$ are possible.}
  \label{fig-es-fey}
\end{figure}

If electrons or quarks are composite, heavy excited fermion states ($f^\ast$)
could be produced in $ep$ reactions at HERA by $t$-channel exchange of photons,
$Z$ or $W$ bosons as shown in the diagram of \fig{es-fey}.  The observation of
such states would be an indication of fermion substructure.

The ZEUS \cite{zeus-tampere-555} and H1 \cite{h1-vancouver-581} collaborations
have searched for excited electrons ($e^\ast$), neutrinos ($\nu^\ast$) and
quarks ($q^\ast$) under the assumption that they decay into standard fermions
and electroweak gauge bosons in one of the modes summarized in \tab{es-decmod}.
The main selection cuts for the $e^\ast$ and $\nu^\ast$ searches required events
consistent with the presence of both a high-$P_t$ final-state lepton (either an
electron identified in the central detector or large missing transverse momentum
indicating a neutrino) and of the gauge boson from the $f^\ast$ decay (either
identified directly in the case of a photon, or via the hadronic or leptonic
decay products in the case of $W$ and $Z$ bosons). H1 also searched for excited
quark states in events with a high-$P_t$ jet and either a photon or an electron
and missing transverse momentum ($\ptmiss$). The backgrounds from various
Standard Model (SM) processes (mainly neutral current (NC) and charged current
(CC) deep-inelastic scattering (DIS), QED Compton scattering (QEDC),
photoproduction of high-$P_t$ jets (PHP) and prompt photons, on-shell production
of $W$ and $Z$ bosons, and lepton-pair production in Bethe-Heitler
($\gamma\gamma$) processes, see \tab{es-decmod}) are estimated from MC
simulations. The signal reactions are simulated using cross sections calculated
according to the model by Hagiwara, Zeppenfeld and Komamiya (HZK)
\cite{hag-85-01} for spin-$1/2$ excited fermions.

\begin{table}
  \caption{Decay modes, event signatures and main SM background sources  
           considered in the $e^\ast$, $\nu^\ast$ and $q^\ast$ searches by ZEUS
           \pcite{zeus-tampere-555} and H1 \pcite{h1-vancouver-581}.}
  \centerline{\small
  \begin{tabular}{llll}
  \topboldline
  $f^\ast$ decay&signature&main SM background&studied by\\
  \midboldline
  $e^\ast\to e+\gamma$ 
    &$e+\gamma$               &QEDC, NC                         & H1, ZEUS\\
  $e^\ast\to e+Z\to e+q\qbar$
    &$e+2$jets                &NC                               & H1, ZEUS\\
  $e^\ast\to e+Z\to e+e^+e^-$
    &$3e$                     &$\gamma\gamma$, NC               & H1\\
  $e^\ast\to e+Z\to e+\nu\nubar$
    &$e+\ptmiss$              &NC, CC, $\gamma\gamma$, $W$      & H1\\
  $e^\ast\to\nu+W\to\nu+q\qbar'$ 
    &$\ptmiss+2$jets          &CC, PHP                          & H1, ZEUS\\
  $e^\ast\to\nu+W\to\nu+e\nu$ 
    &$e+\ptmiss$              &NC, CC, $\gamma\gamma$, $W$      & H1\\
  \hline
  $\nu^\ast\to\nu+\gamma$ 
    &$\gamma+\ptmiss$         &CC, NC                           & H1, ZEUS\\
  $\nu^\ast\to e+W\to e+q\qbar'$
    &$e+2$jets                &NC                               & H1\\
  $\nu^\ast\to e+W\to e+e\nu$
    &$2e+\ptmiss$             &$\gamma\gamma$, QEDC, $W$        & H1\\
  $\nu^\ast\to\nu+Z\to\nu+q\qbar$ 
    &$\ptmiss+2$jets          &CC, PHP                          & H1\\
  $\nu^\ast\to\nu+Z\to\nu+e^+e^-$ 
    &$2e+\ptmiss$             &$\gamma\gamma$, QEDC, $W$        & H1\\
  \hline
  $q^\ast\to q+\gamma$ 
    &$\gamma+$jet             &prompt $\gamma$                  & H1\\
  $q^\ast\to q'+W\to q'+e\nu$
    &$e+\ptmiss+$jet          &NC, CC, $W$                      & H1\\
  \bottomboldline
  \end{tabular}}
  \label{tab-es-decmod}
\end{table}

\begin{figure}[b]
  \sidecaption
  \psfig{file=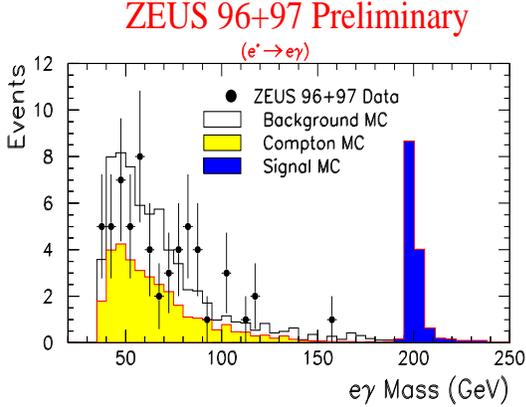,width=6.9cm,height=5.4cm}\kern-2.mm
  \caption{Spectrum of the invariant $e\gamma$ mass of ZEUS candidates for 
           $e^\ast$ production and subsequent decay $e^\ast\to e\gamma$. The
           points with error bars represent the data, the open (light-shaded)
           histogram the total SM background (the contribution from QED Compton
           scattering). The dark-shaded histogram illustrates the signal
           expected for a hypothetical $e^\ast$ with a mass of $200\gev$.}
  \label{fig-es-egam}
\end{figure}

\begin{figure}[t]
  \centerline{\strut\kern5.mm
  \psfig{file=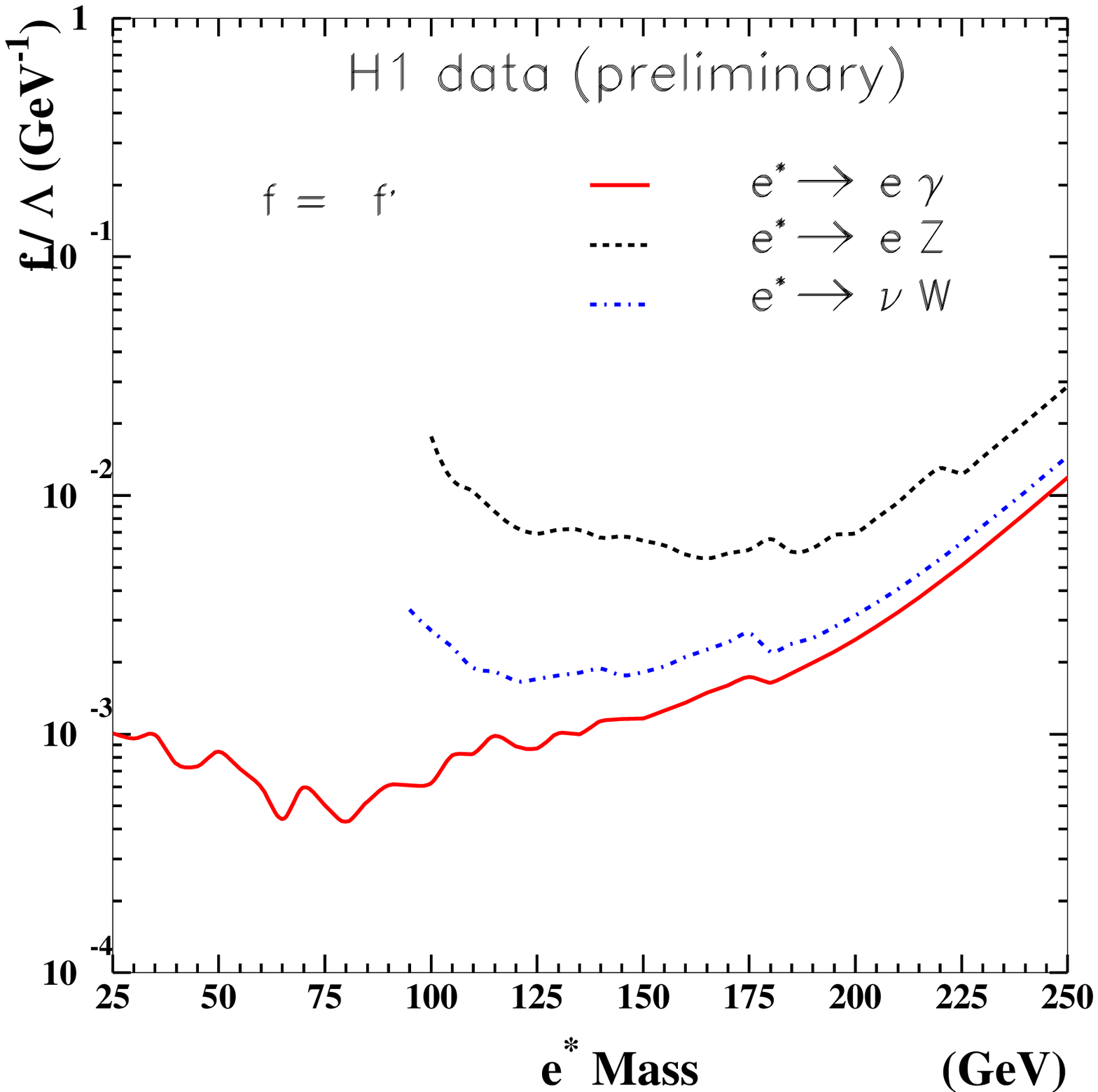,width=5.6cm,height=5.55cm}
  \psfig{file=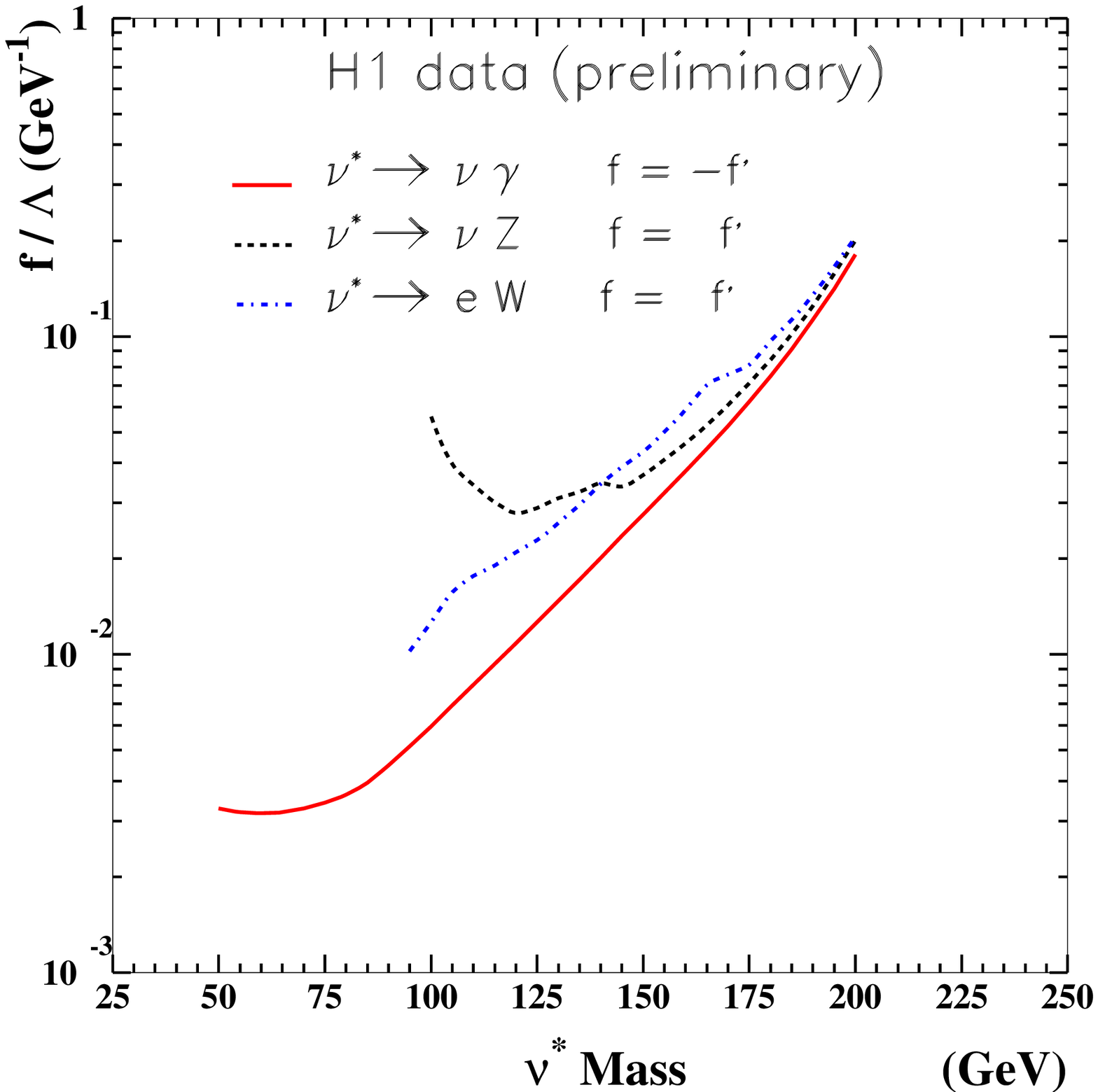,width=5.6cm,height=5.55cm}}
  \centerline{
  \psfig{file=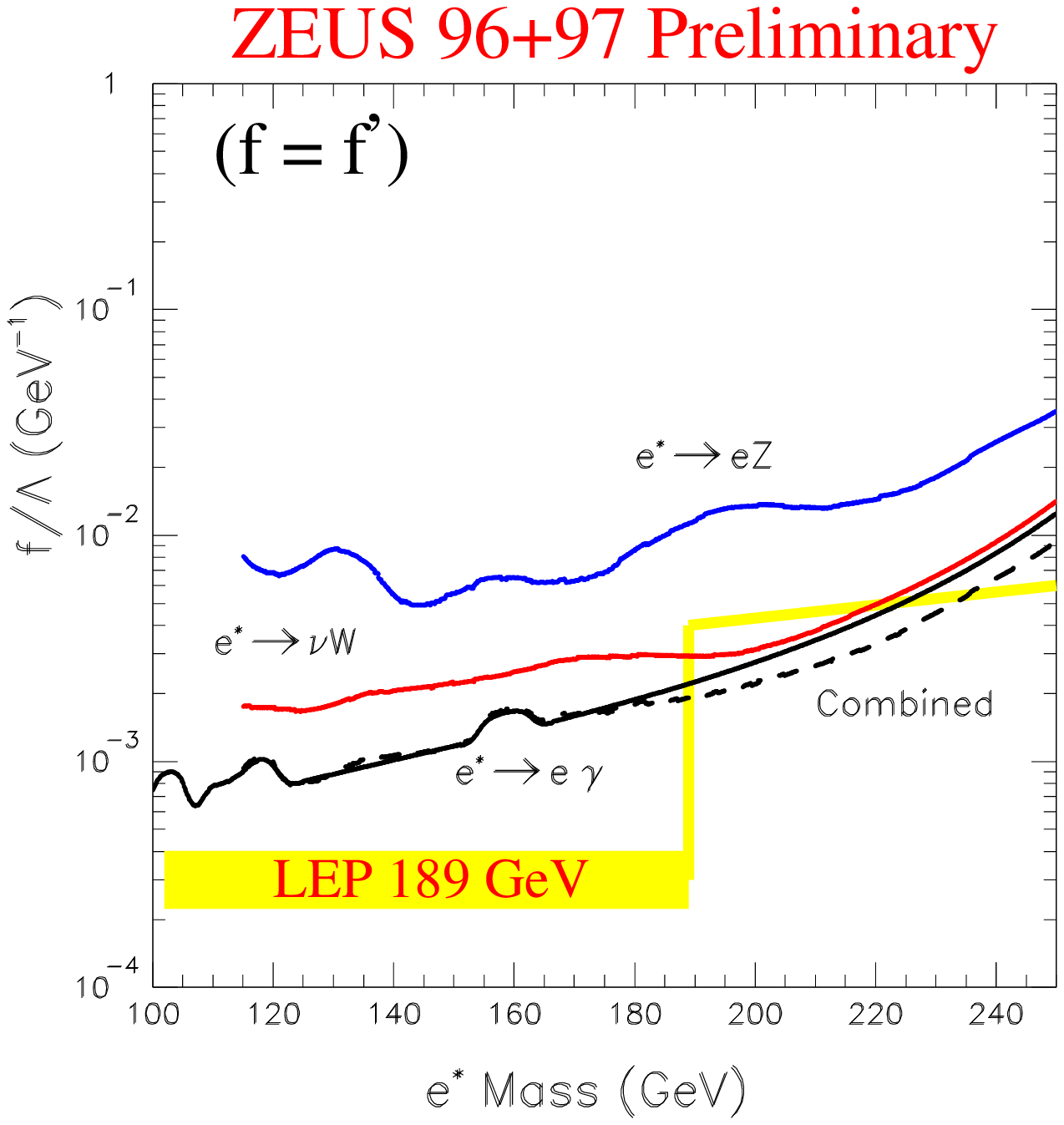,width=5.6cm,height=6.0cm}
  \psfig{file=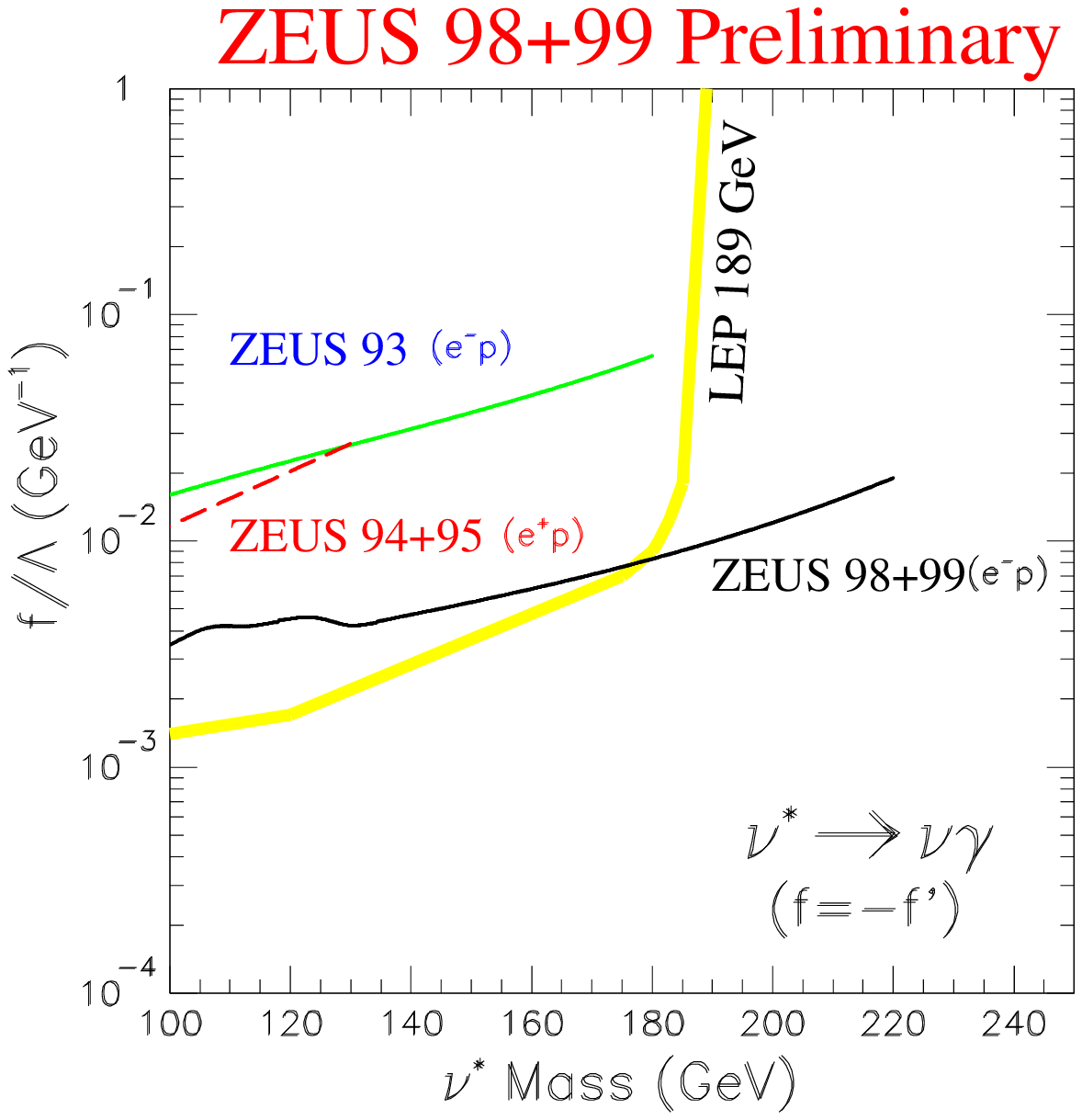,width=5.6cm,height=6.0cm}}
  \caption{Limits at $95\%$ C.L.\ on $f/\Lambda$ from H1 (top) and ZEUS (bottom)
           searches for excited electrons (left) and excited neutrinos
           (right). The shaded bands denoted ``LEP $189\gev$'' indicate the
           exclusion limits from \pcite{and-99-01}, with the band width
           approximating the amplitude of their small-scale variations.}
  \label{fig-es-flam}
\end{figure}

No indication of a signal above SM background has been found. As an example,
\fig{es-egam} shows the distribution of the invariant $e\gamma$ mass in the
event sample selected in the ZEUS search for $e^\ast\to e\gamma$ decays. For
$f^\ast$ masses $M_{f^\ast}\gtrsim100\gev$ the searches are almost
background-free, yielding upper limits at $95\%$ C.L.\ on the cross section
times branching ratio ($\sigma\cdot\BR$) of typically $0.1$--$1\pb$. The HZK
model is used to relate these limits to exclusion plots of $f/\Lambda$ vs.\
$M_{f^\ast}$, where $\Lambda$ is the characteristic mass scale and $f$
determines the coupling at the $f^\ast$-fermion-boson vertex. In the HZK model,
there are three independent couplings, of which the one to gluons ($f_s$) is
irrelevant for $f^\ast$ production at HERA and those to the U(1) and SU(2) gauge
fields are assumed to be related by $f\equiv|f'_{U(1)}|=|f_{SU(2)}|$. The HERA
limits shown in \fig{es-flam} are by almost one order of magnitude stronger than
those based on the 1993--1995 data \cite{h1-97-01,zeus-97-08} and exclude
$f/\Lambda$ values above $\ord{10^{-3}\dots10^{-2}\gev^{-1}}$ for $f^\ast$
masses between $25$ and $220\gev$; this also applies to the H1 results on
$q^\ast$ (not shown). Assuming $f/\Lambda=1/M_{f^\ast}$, the results exclude the
existence of $e^\ast$ with masses between $25$ and $229\gev$, and of $\nu^\ast$
between $25$ and $161\gev$. The $\nu^\ast$ limits are dominated by the recent
$e^-p$ results (from ZEUS) since the expected $\nu^\ast$ cross section is more
than an order of magnitude larger for $e^-p$ than for $e^+p$ scattering.  The
comparison with the latest LEP limits \cite{and-99-01} in \fig{es-flam}
demonstrates that the HERA searches, in particular in the $\nu^\ast$ channel,
advance into new discovery windows which are not (yet) accessible to other
experiments. The H1 limits for $q^\ast$ (assuming $f_s\eql0$) complement those
from Tevatron where $q^\ast$ production requires non-zero $f_s$.

\section{Leptoquarks}
\label{sec-lq}

\begin{figure}[ht]
  \sidecaption
  \begin{picture}(125.,70.)(0.,0.)
  \SetOffset(0.,-30.)
  \ArrowLine(10,30)(40,65)           
  \ArrowLine(10,100)(40,65)          
  \Line(40,65)(80,65)                
  \ArrowLine(80,65)(110,30)          
  \ArrowLine(80,65)(110,100)         
  \Text(10,90)[t]{$e$}              
  \Text(110,90)[lt]{$\ell_i$}       
  \Text(10,40)[b]{$q_j$}             
  \Text(110,40)[lb]{$q_k$}           
  \Text(60,70)[b]{LQ}                
  \Text(30,65)[rc]{$\lambda_{1j}$}   
  \Text(90,65)[lc]{$\lambda_{ik}$}   
  \end{picture}
  \caption{Feynman graph of the resonant production of a $F\eql2$ LQ in
           electron-quark scattering. The subscripts of the Yukawa couplings
           at production and decay vertex of the LQ indicate the generations
           of the leptons and quarks involved.}
  \label{fig-lq-fey}
\end{figure}
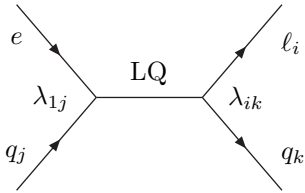

Leptoquarks (LQ's) are hypothesized bosons coupling to lepton-quark pairs.  They
could be resonantly produced in $ep$ reactions according to the diagram shown in
\fig{lq-fey}.  Buchm\"uller, R\"uckl and Wyler (BRW) \cite{buc-87-01} have
classified LQ's which (i) conserve the SM gauge symmetries, (ii) only couple to
quarks, leptons and SM gauge bosons and (iii) only have flavor-diagonal
couplings, in ten different states characterized by their fermion number
($|F|\eql2$ or $F\eql0$), spin (scalar ($S$) or vector ($V$)) and weak isospin.
Premise (ii) implies that LQ's only decay into $\ell^\pm q$ or $\nu q'$ pairs
with given branching ratios. In many recent LQ searches, this assumption is
dropped and a variable branching ratio $\beta_\ell=\BR(\LQ\to\ell^\pm q)$ is
assigned to the LQ's (accounting e.g.\ for the case of $R_P$-violating squarks
\cite{lem-99-01}).

\subsection{First-generation leptoquarks}
\label{sec-lq-fg}

LQ's coupling only to first-generation leptons can be produced at HERA in
reactions of the type $ep\to\LQ+X\to eq+X\;(\nu_e q'+X)$, yielding the same
event topologies as NC (CC) DIS at high $eq$ invariant mass. The distinctive LQ
signature is a resonant distribution of $\sqrt{xs}$ around the LQ mass,
$M_\LQsub$, where $\sqrt{s}$ is the $ep$ center-of-mass energy and the Bjorken
variable $x$ is the fraction of the proton momentum carried by the struck quark.
Furthermore, the LQ decay angular distribution implies harder distributions of
$y=s/(xQ^2)$ than for DIS ($Q^2$ is the negative square of the four-momentum
transfer between $e$ and $p$). LQ production at HERA has received particular
attention after H1 \cite{h1-97-06} and ZEUS \cite{zeus-97-03} have reported an
excess of events of NC DIS at high $x$ and $Q^2$ in their 1994--1996 data.

\begin{figure}[t]
  \centerline{
  \psfig{file=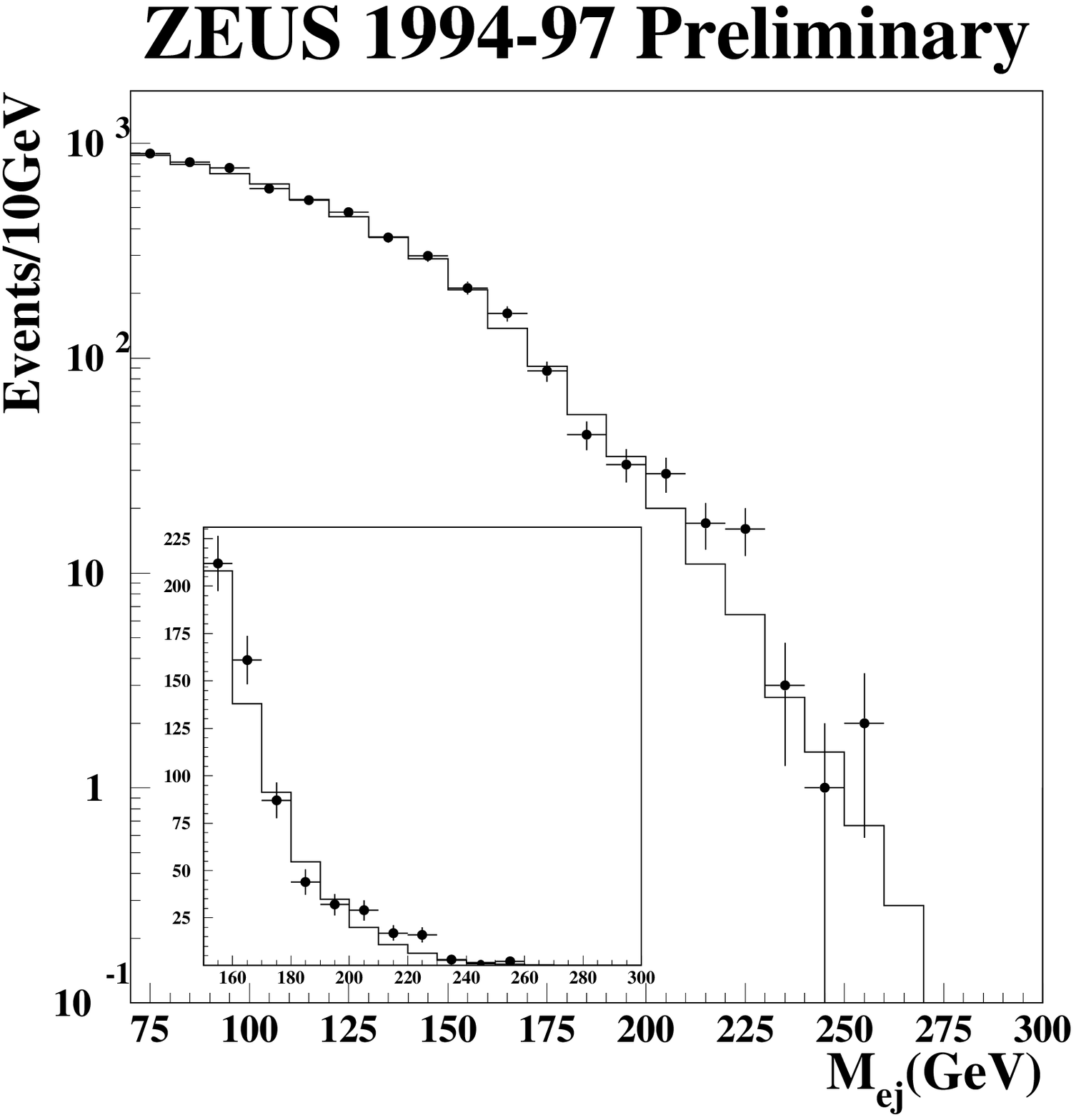,width=5.6cm,height=5.cm}
  \psfig{file=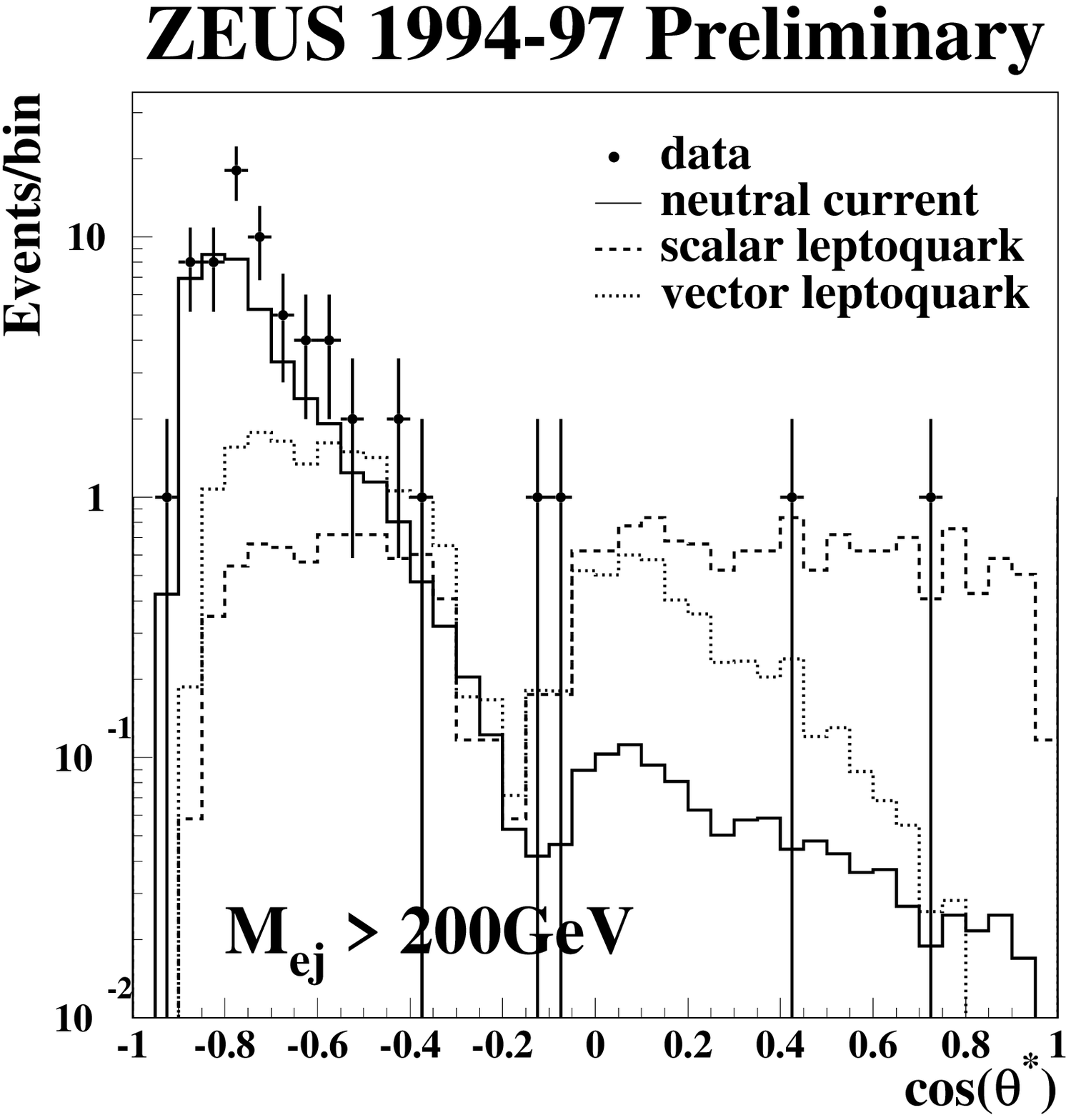,width=5.6cm,height=5.cm}}
  \sidecaption
  \psfig{file=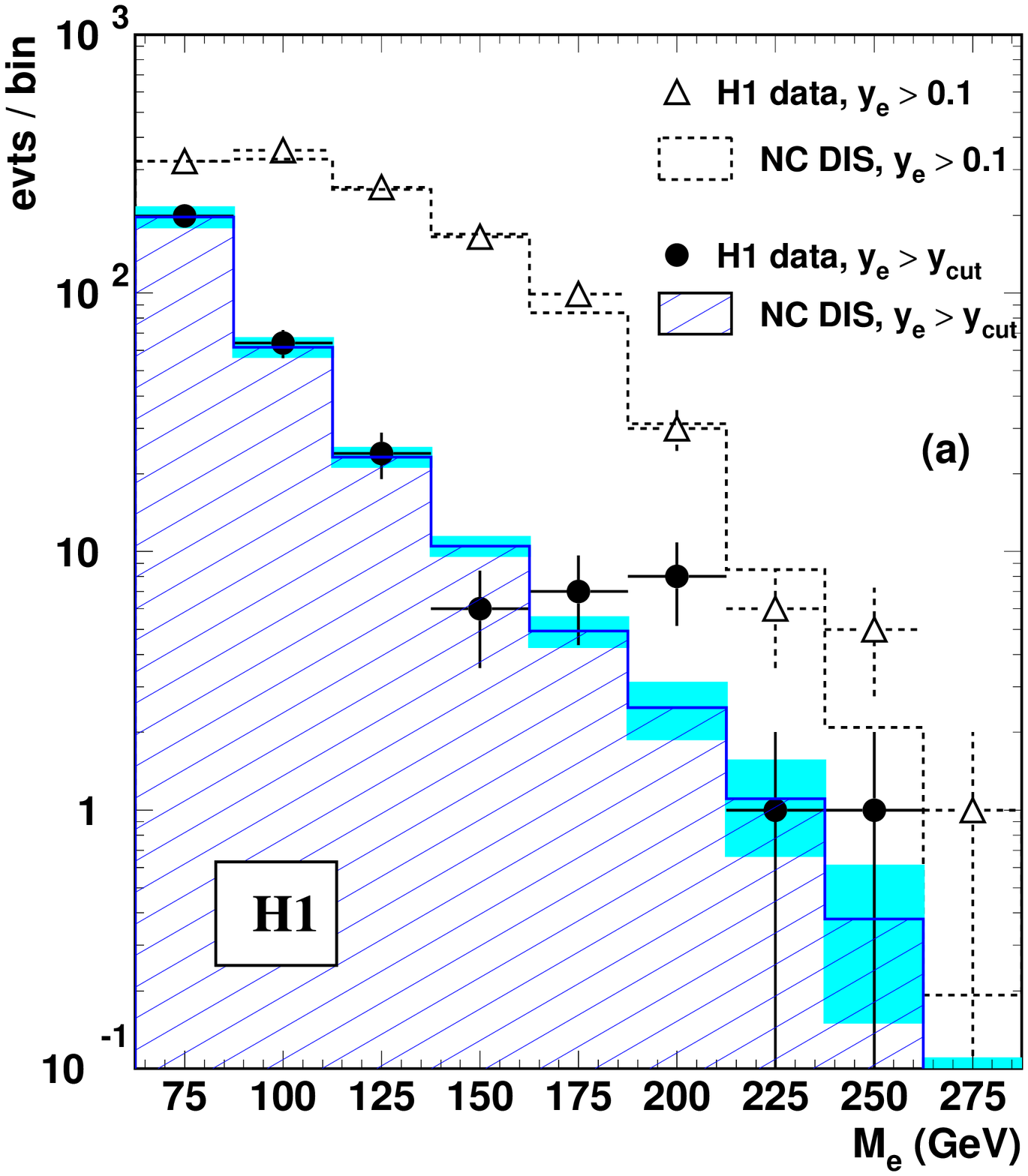,width=5.6cm,height=5.cm}
  \caption{Distributions of $M_{ej}$ (top left, ZEUS) and $M_e$ (bottom, H1) in
           the $e^+p$ NC data. The symbols with statistical error bars represent
           the data, the histograms the DIS MC.  The H1 data are shown before
           (open triangles) and after (full dots, hatched histogram) a cut
           $y\gre y_\smin(M_e)$. The shaded area indicates the uncertainty of
           the SM prediction. The polar angle distribution in the $eq$
           center-of-mass system is shown in the top right plot for ZEUS events
           with $M_{ej}\gre200\gev$.}
  \label{fig-lq-mdis}
\end{figure}

ZEUS \cite{zeus-vancouver-754,zeus-tampere-546,zeus-tampere-552} and H1
\cite{h1-99-03} have searched in the available DIS data for deviations from the
SM prediction consistent with a LQ signal. The distributions of
$M_e=\sqrt{x_es}$ (derived from energy and angle of the scattered electron, H1)
and $M_{ej}$ (invariant $e$-jet mass, ZEUS) are shown in \fig{lq-mdis}. The
excess in the H1 data is still present at $M_e\approx200\gev$ but has not been
corroborated by the 1997 data. Also ZEUS observes an excess at $M_{ej}\gtrsim
200\gev$; however, the decay angular distribution does not support a LQ
interpretation (see \fig{lq-mdis}). No significant deviations from the SM
predictions are observed in CC DIS \cite{zeus-tampere-546,h1-99-03} and
in the new $e^-p$ data \cite{zeus-tampere-552}.

\begin{figure}[t]
  \centerline{
  \begin{minipage}{5.5cm}
  \centerline{
  \psfig{file=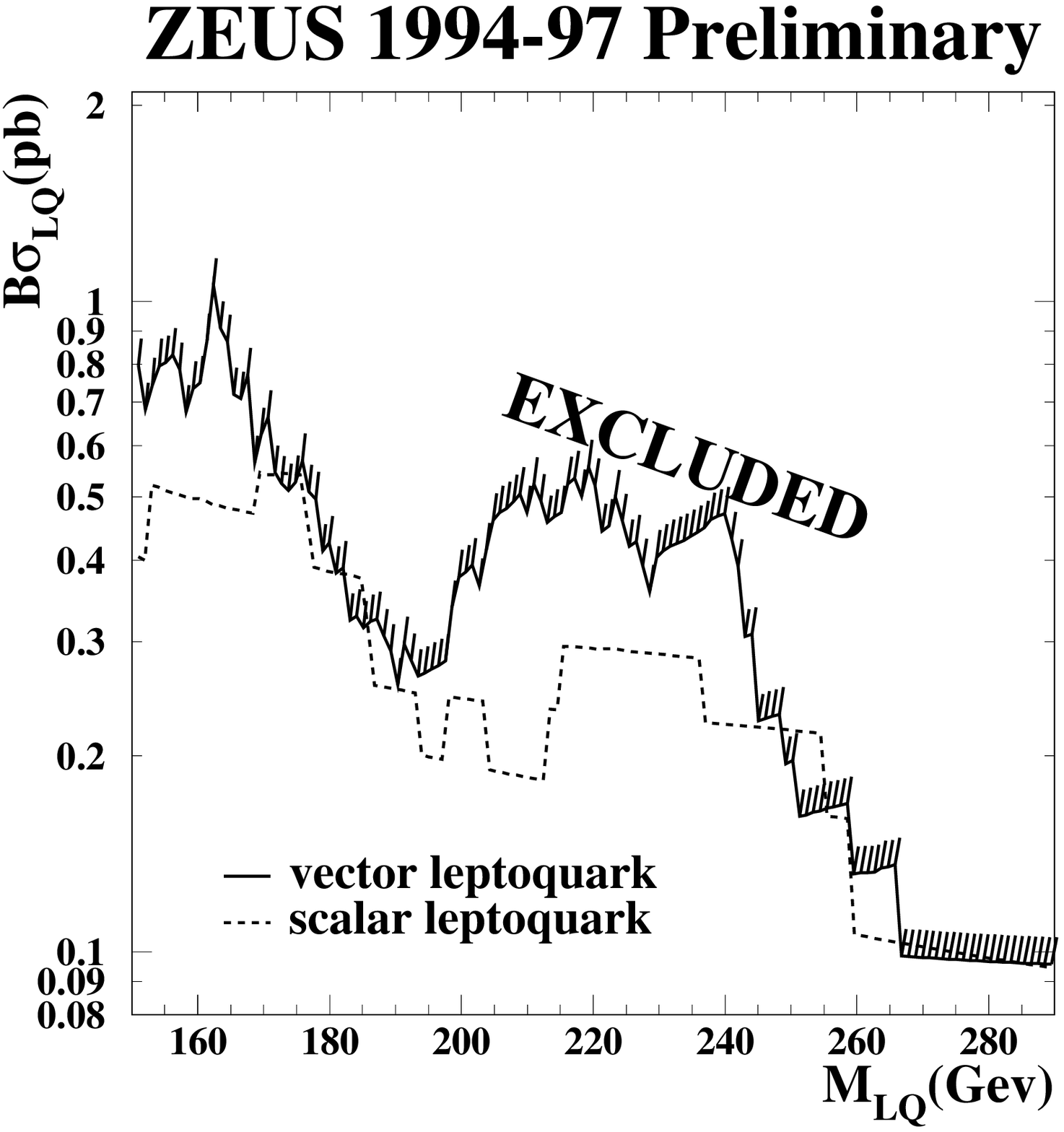,width=5.5cm,height=4.6cm}}
  \centerline{
  \psfig{file=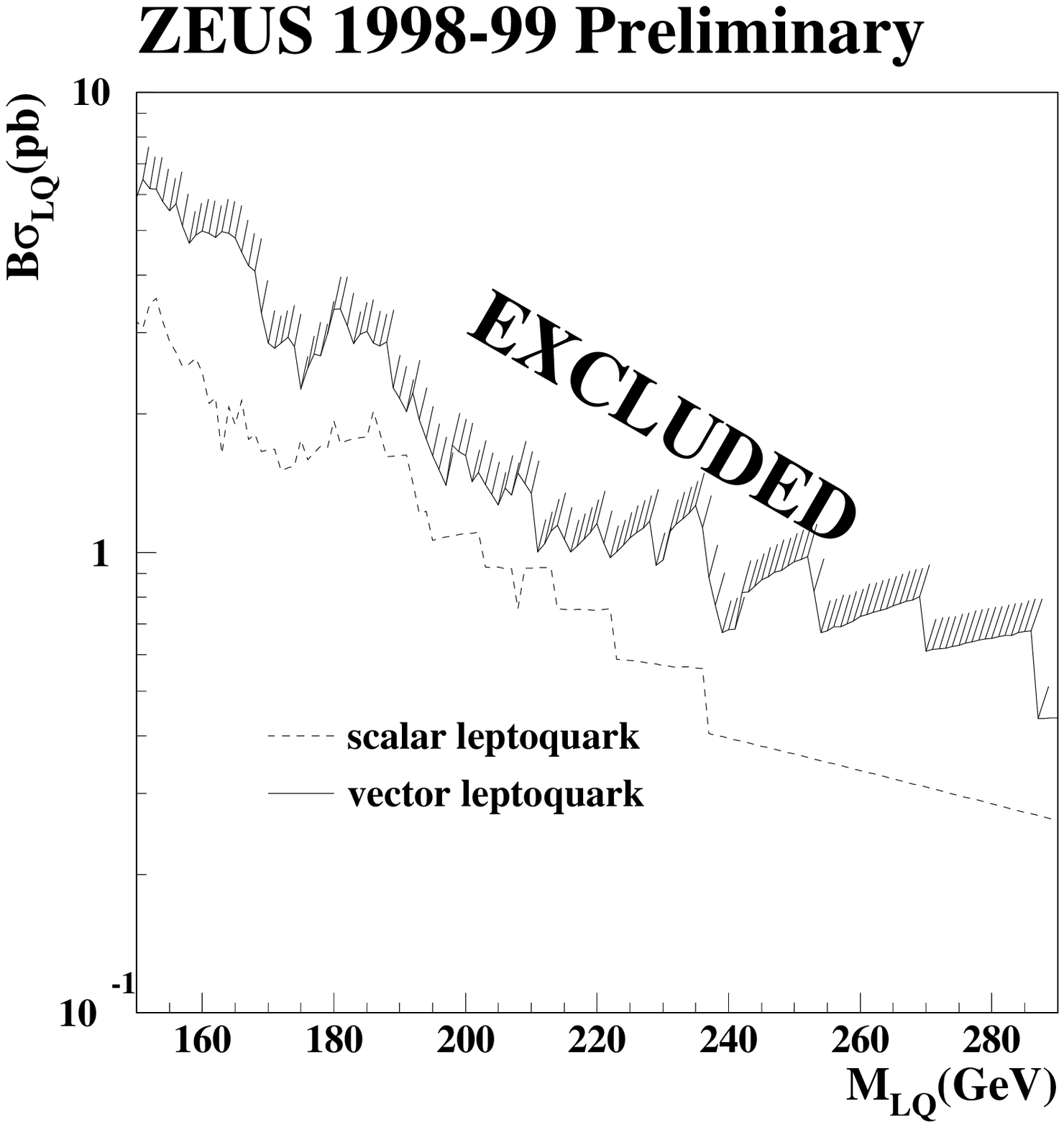,width=5.5cm,height=4.6cm}}
  \caption{ZEUS limits at $95\%$ C.L.\ on $\sigma\cdot\beta$ for scalar (dotted
           lines) and vector (full lines) LQ's, from the $e^+p$ 
           \pcite{zeus-vancouver-754} (top) and the $e^-p$
           \pcite{zeus-tampere-552} (bottom) data.}
  \label{fig-lq-zeussb}
  \end{minipage}
  \begin{minipage}{5.5cm}
  \centerline{
  \psfig{file=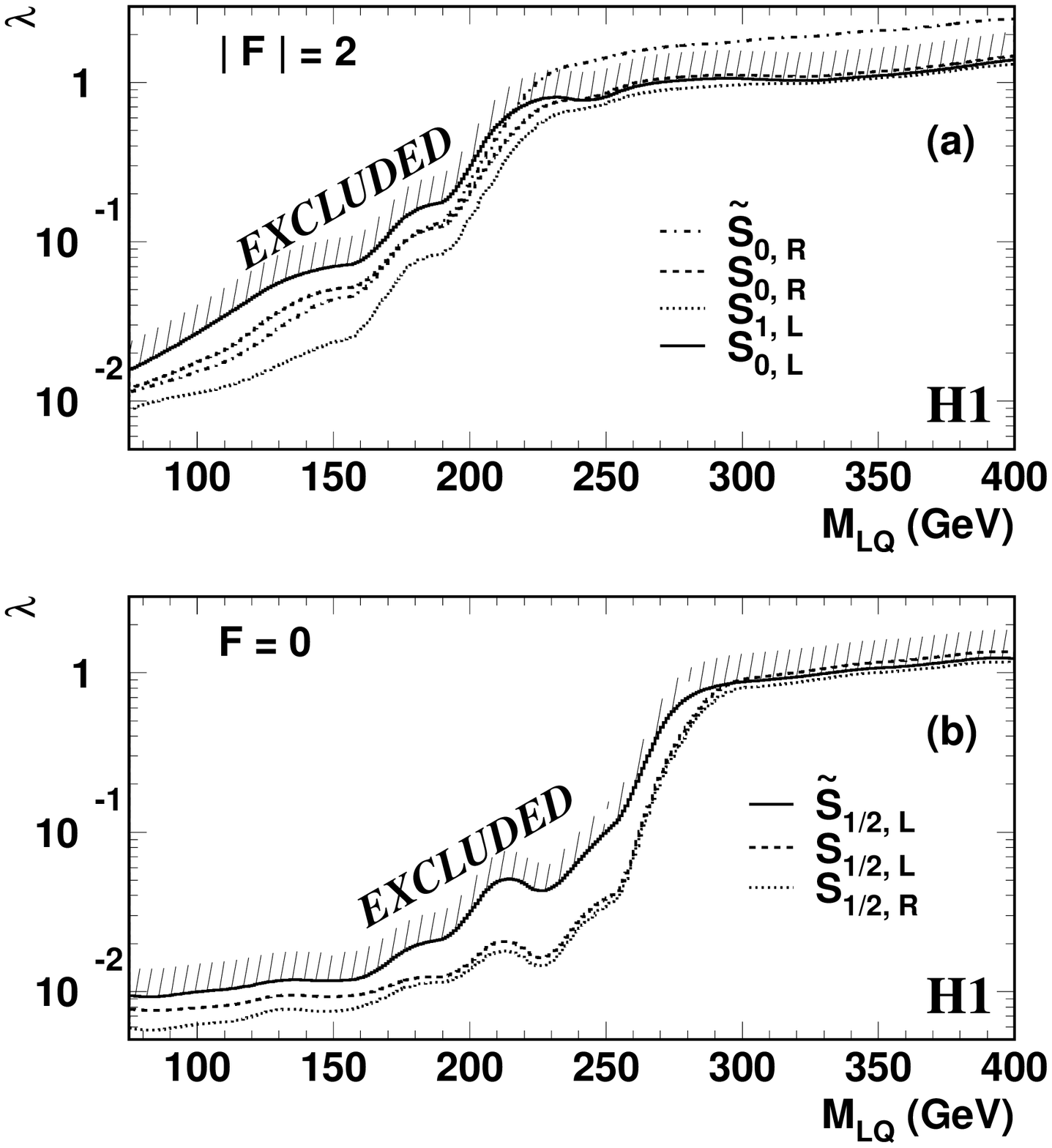,width=5.5cm,height=9.27cm}}
  \caption{H1 limits at $95\%$ C.L.\ on $\lambda=\lambda_{1j}$ for different
           scalar LQ's with $|F|\eql2$ (top) and $F\eql0$ (bottom). Only NC-type
           data are used for these limits.}
  \label{fig-lq-h1lam}
  \end{minipage}}
  \vskip3.mm
  \sidecaption
  \psfig{file=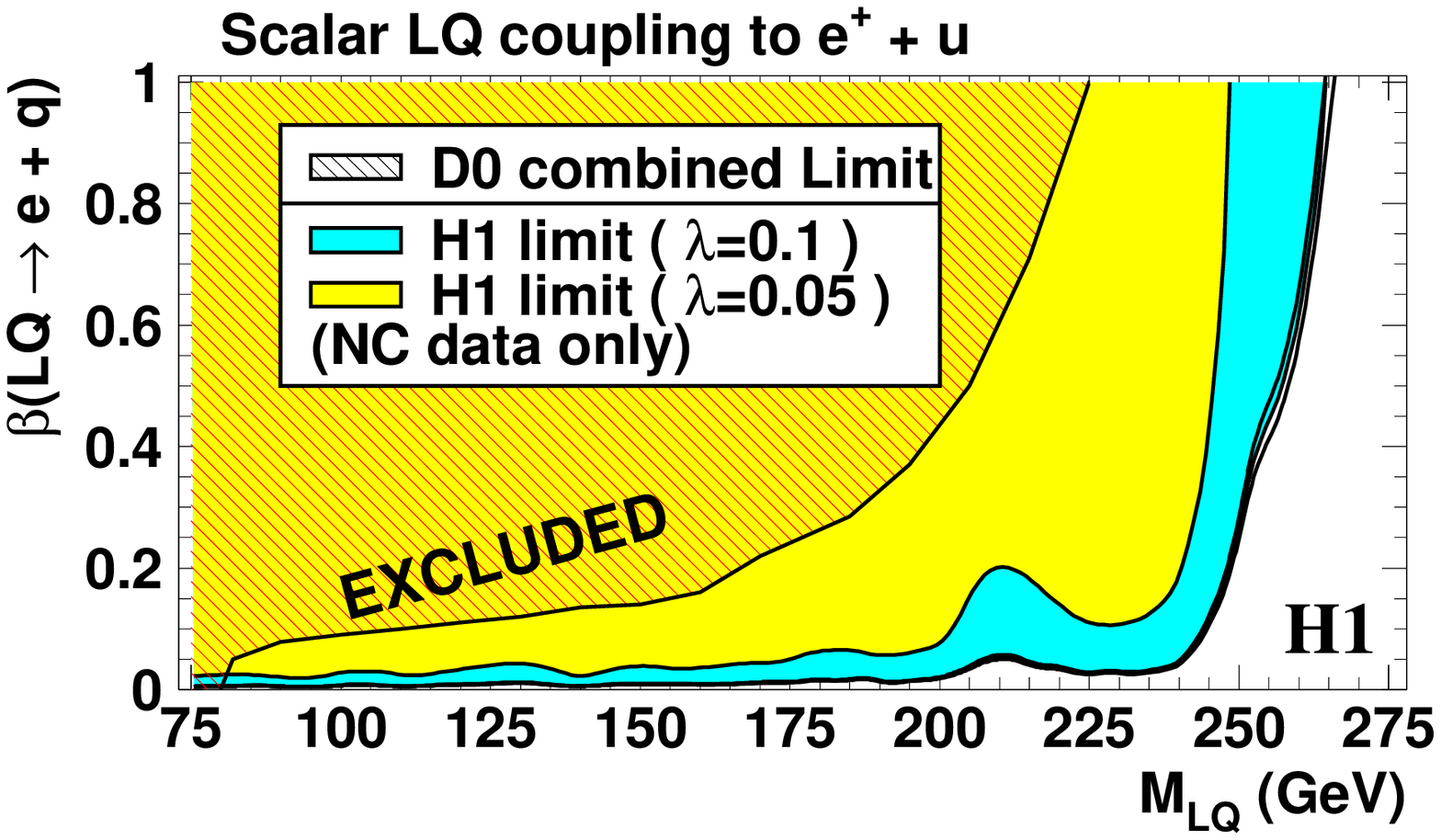,width=7.1cm,height=4.3cm}
  \caption{H1 exclusion limits on $\beta_e$ as functions of $M_\LQsub$, for
           scalar $F\eql0$ LQ's coupling to $u$ quarks, for different fixed
           values of $\lambda_{1j}$. The cross-hatched area indicates the region
           excluded by \DO \pcite{abb-97-01}. Only NC-type data have been used
           for this plot.}
  \label{fig-lq-h1beta}
\end{figure}

Using a MC simulation of signal events according to the BRW cross sections,
upper limits on $\sigma\cdot\BR$ for LQ production have been set. The ZEUS
exclusion plots for NC-type LQ decays in $e^+p$ and $e^-p$ scattering are shown
in \fig{lq-zeussb}; the H1 $e^+p$ results (not shown) are similar but extend to
higher $M_\LQsub$ since $u$-channel LQ exchange and DIS-LQ interference are
taken into account in addition to $s$-channel LQ formation.  Assuming fixed
$\beta_e$ according to BRW, upper limits on the coupling $\lambda\eql
\lambda_{1j}$ are derived as functions of $M_\LQ$ for different LQ types
(\fig{lq-h1lam} shows the H1 result for scalar LQ's).  Note that the $e^+p$ data
are much more sensitive to $F\eql0$ LQ's, which can be produced from valence
quarks, than to $|F|\eql2$ LQ's. $M_\LQ$ limits which are independent of
$\lambda$ are obtained by the Tevatron experiments from searches for LQ pair
production in $p\pbar$ reactions. The combined CDF and \DO limit is
$M_\LQ\gre242\gev$ \cite{gro-98-01} for scalar LQ's with $\beta_e\eql1$ and is
expected to be even higher for vector LQ's. H1 has presented the LQ limits as
functions of $\beta_e$ and $M_\LQ$ for fixed values of $\lambda_{1j}$
(\fig{lq-h1beta}), thus allowing a direct comparison to the LQ exclusion region
from \DO \cite{abb-97-01}. The HERA experiments have a unique discovery
potential at low $\beta_e$ and high $M_\LQsub$; future high-statistics HERA data
is needed in order to clarify whether the anomalies observed in the $e^+p$ data
are an indication of LQ production.

\subsection{Lepton-flavor violating leptoquarks}
\label{sec-lq-lfv}

\begin{figure}[t]
  \centerline{
  \begin{minipage}{5.5cm}
  \centerline{
  \psfig{file=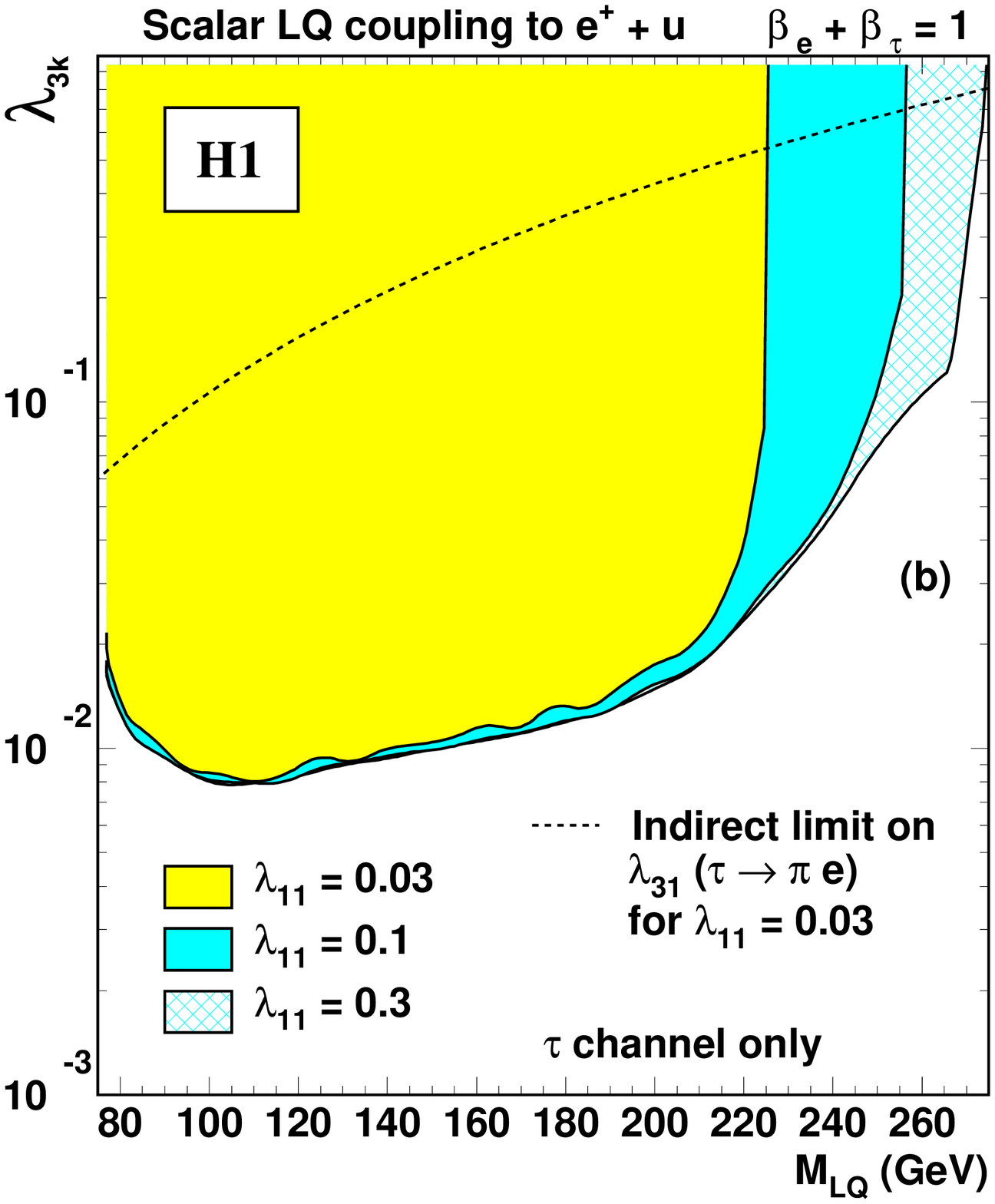,width=5.5cm,height=5.5cm}}
  \caption{H1 exclusion limits at $95\%$ C.L.\ on $\lambda_{3k}$ for different
           fixed values of $\lambda_{1j}$. Scalar LQ's which only couple to $e$
           and $\tau$ ($\beta_e+\beta_\tau=1$) are assumed. Also indicated is an
           indirect limit derived from the non-observation of the lepton-flavor
           violating decay $\tau\to e\pi$.}
  \label{fig-lq-h1lam3}
  \end{minipage}
  \hfill
  \begin{minipage}{5.5cm}
  \centerline{
  \psfig{file=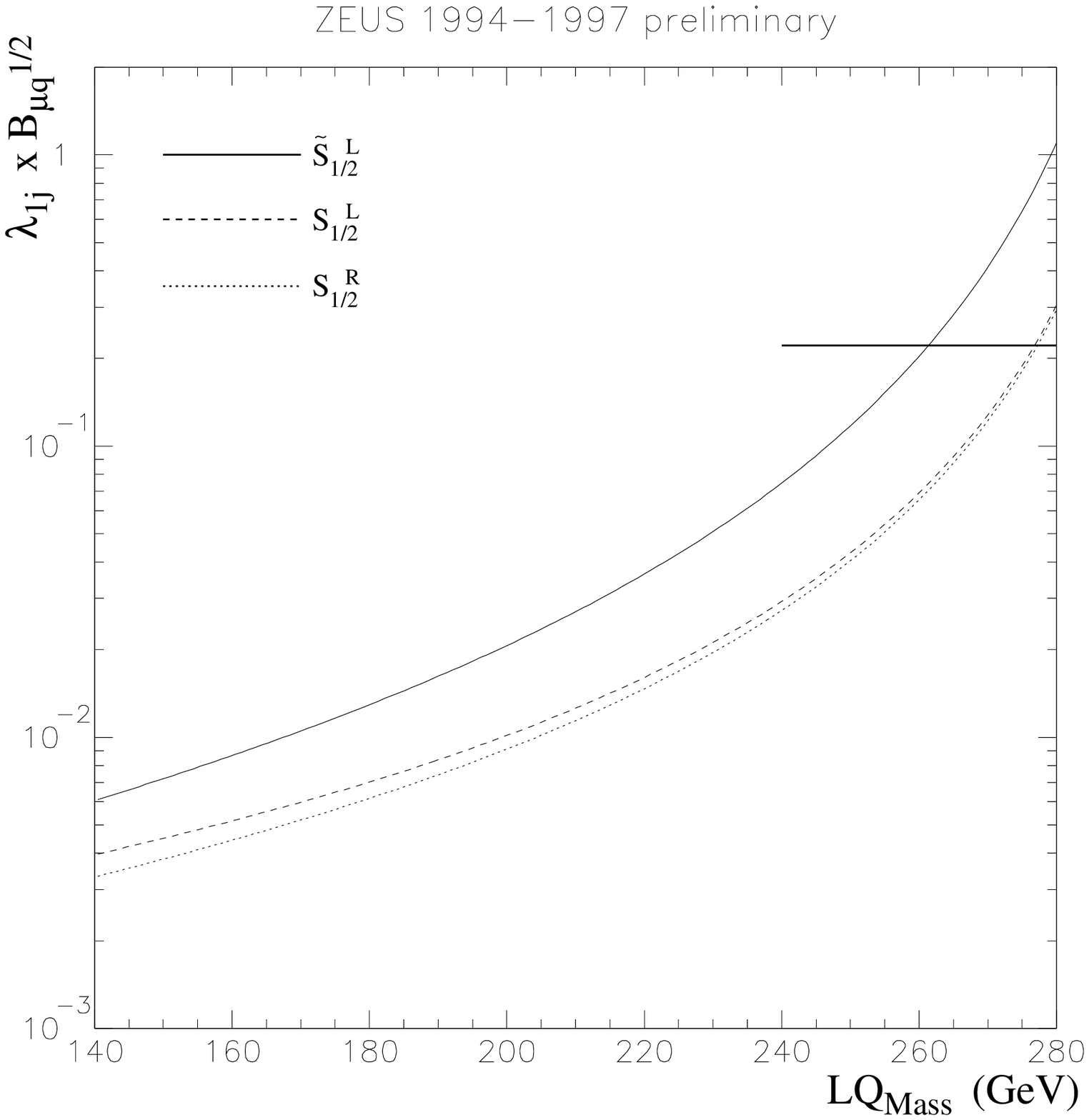,width=5.5cm,height=5.5cm}}
  \caption{ZEUS exclusion limits at $95\%$ C.L.\ on $\lambda_{1j}\cdot\BR(LQ\to
           \mu q)$ for three types of scalar LQ's coupling to electrons and
           muons and to $d$, $u$, and $d$ and $u$ quarks, respectively (top to
           bottom). The horizontal line indicates electromagnetic coupling
           strength.}
  \label{fig-lq-zeuslfv}
  \end{minipage}}
\end{figure}

H1 has searched for instances of the reaction $ep\to\LQ+X\to\tau q+X$, with
subsequent hadronic $\tau$ decays \cite{h1-99-03}. The event selection requires
the presence of a narrow hadronic jet with small track multiplicity, which is
azimuthally opposite to the direction of the net transverse momentum that has to
exceed $10\gev$. No events fulfilling these criteria have been observed,
allowing to set limits on the LQ-$\tau$ coupling $\lambda_{3k}$ for fixed values
of $\lambda_{1j}$ (\fig{lq-h1lam3}).  These limits are stronger than those
derived from lepton-flavor violating decays of $\tau$'s. The limits from
searches for LQ's decaying to $\tau$'s at Tevatron are $M_\LQsub>99\gev$ (CDF
\cite{abe-96-01}) and $>94\gev$ (\DO \cite{abb-98-06}).

ZEUS has performed a search for lepton flavor violation in the $\mu$ channel,
$ep\to\LQ+X\to\mu q+X$ \cite{zeus-tampere-551}. The selection requires event
topologies similar to high-$Q^2$ NC DIS reactions, except that the scattered $e$
is replaced by a $\mu$. No such event is found, with an expected background of
$0.3$ events predominantly from Bethe-Heitler muon pair production. Upper limits
are set on $\lambda_{1j}\cdot\BR(LQ\to\mu q)$ (\fig{lq-zeuslfv}). Assuming
couplings of electromagnetic strength, $\lambda_{1j}=\lambda_{3k}
=\sqrt{4\pi\alpha}$, these limits correspond to $M_\LQsub>262$--$285\gev$,
depending on the LQ type.

\section{Contact interactions}
\label{sec-ci}

\begin{table}[t]
  \renewcommand{\arraystretch}{0.938}
  \newcolumntype{n}{@{\kern3.5pt}c@{\kern3.5pt}}
  \newcolumntype{o}{c@{\kern3.5pt}}
  \newcolumntype{p}{@{\kern3.5pt}c}
  \newcolumntype{q}{@{\kern5.pt}c@{\kern5.pt}}
  \caption{CI combinations considered in the ZEUS \pcite{zeus-99-05} and H1 
           \pcite{h1-tampere-157f} analyses, and the corresponding $95\%$ C.L.\
           lower limits on $\Lambda$ derived from the NC DIS data.}
  \centerline{
  \begin{tabular}{lonnnnnnprrrr}
  \topboldline
    &&&&&&&&&\multicolumn{4}{c}{\ \ $\Lambda$ limit $[\Tev]$}\\
    CI&
    $\eta_\LL^u$&$\eta_\LR^u$&$\eta_\RL^u$&$\eta_\RR^u$&
    $\eta_\LL^d$&$\eta_\LR^d$&$\eta_\RL^d$&$\eta_\RR^d$&
    \multicolumn{2}{c}{\ \ ZEUS}&\multicolumn{2}{c}{H1}\\
    &&&&&&&&&$\epsilon=-1$&$+1$
            &$         -1$&$+1$\\
  \midboldline
    VV&$+$&$+$&$+$&$+$&$+$&$+$&$+$&$+$& 5.0 & 4.7 & 1.9 & 5.0 \\
    AA&$+$&$-$&$-$&$+$&$+$&$-$&$-$&$+$& 3.7 & 2.6 & 3.4 & 2.0 \\
    VA&$+$&$-$&$+$&$-$&$+$&$-$&$+$&$-$& 2.6 & 2.5 & 2.6 & 2.6 \\[3.pt]
    X1&$+$&$-$&$0$&$0$&$+$&$-$&$0$&$0$& 2.8 & 1.8 &\none&\none\\
    X2&$+$&$0$&$+$&$0$&$+$&$0$&$+$&$0$& 3.1 & 3.4 &\none&\none\\
    X3&$+$&$0$&$0$&$+$&$+$&$0$&$0$&$+$& 2.8 & 2.9 & 1.3 & 3.1 \\
    X4&$0$&$+$&$+$&$0$&$0$&$+$&$+$&$0$& 4.3 & 4.0 & 1.8 & 4.1 \\
    X5&$0$&$+$&$0$&$+$&$0$&$+$&$0$&$+$& 3.3 & 3.5 &\none&\none\\
    X6&$0$&$0$&$+$&$-$&$0$&$0$&$+$&$-$& 1.7 & 2.8 &\none&\none\\[3.pt]
    U1&$+$&$-$&$0$&$0$&$0$&$0$&$0$&$0$& 2.6 & 2.0 &\none&\none\\
    U2&$+$&$0$&$+$&$0$&$0$&$0$&$0$&$0$& 3.9 & 4.0 &\none&\none\\
    U3&$+$&$0$&$0$&$+$&$0$&$0$&$0$&$0$& 3.5 & 3.7 &\none&\none\\
    U4&$0$&$+$&$+$&$0$&$0$&$0$&$0$&$0$& 4.8 & 4.4 &\none&\none\\
    U5&$0$&$+$&$0$&$+$&$0$&$0$&$0$&$0$& 4.2 & 4.0 &\none&\none\\
    U6&$0$&$0$&$+$&$-$&$0$&$0$&$0$&$0$& 1.8 & 2.4 &\none&\none\\[3.pt]
    LL&$+$&$0$&$0$&$0$&$+$&$0$&$0$&$0$&\none&\none& 1.2 & 2.3 \\
    LR&$0$&$+$&$0$&$0$&$0$&$+$&$0$&$0$&\none&\none& 1.5 & 3.0 \\
    RL&$0$&$0$&$+$&$0$&$0$&$0$&$+$&$0$&\none&\none& 1.5 & 3.0 \\
    RR&$0$&$0$&$0$&$+$&$0$&$0$&$0$&$+$&\none&\none& 1.2 & 2.3 \\
  \bottomboldline
  \end{tabular}}
  \label{tab-ci-sum}
\end{table}

Contact interactions (CI) provide an effective phenomenology to describe the
cross section modification of $ep$ NC DIS due to new physics at mass scales far
beyond the HERA center-of-mass energy, e.g.\ $s$- and $u$-channel exchange of
LQ's or exchange interactions in presence of a common substructure of electrons
and quarks. The Lagrangian of the vector $eeqq$ CI investigated at HERA is
characterized by an overall strength $\epsilon g^2/\Lambda^2$ (where $g^2=4\pi$
by convention, $\Lambda$ is the effective mass scale and $\epsilon$ is an
overall sign) and by a set of chiral couplings $\eta_{ab}^q$, where $q=u,d$ and
$a,b=L,R$ denote the handedness of the couplings to leptons and quarks.
Depending on $\epsilon$, the CI-SM interference ($\propto Q^2/\Lambda^2$)
enhances or decreases the cross section at intermediate $Q^2$, whereas the pure
CI part ($\propto Q^4/\Lambda^4$) increases the cross section at highest
$Q^2$. No indications for the presence of CI have been found in correspondig
searches by ZEUS \cite{zeus-99-05} and H1 \cite{h1-tampere-157f}.  The CI
patterns investigated and the resulting $95\%$ C.L.\ limits on $\Lambda$ are
summarized in \tab{ci-sum}. Some of the CI scenarios are investigated for the
first time, for others the limits reported by the LEP and Tevatron experiments
are mostly stronger than those in \tab{ci-sum} (see \cite{zeus-99-05} and
references therein). Limits of $\ord{10\tev}$ can furthermore be derived from
atomic parity violation measurements (APV) \cite{cas-99-01,ros-99-01} for
parity-violating CI scenarios (the last four rows in \tab{ci-sum}). H1 has
converted the appropriate $u$ and $d$ combinations of the purely chiral limits
(LL etc.) into limits on $M_\LQsub/\lambda_{1j}$ for different LQ species,
ranging from $\approx 200\gev$ to almost $1\tev$ (valid for $M_\LQsub\gg\sqrt
s$); however, it should be noted that stronger limits are obtained from APV.

\section*{Acknowledgments} 
None of the research reported here would have been possible without the
dedicated efforts of the HERA crew, and of all those who contributed to design,
construction, maintenance and operation of the detectors. I'm indebted to
Emmanuelle Perez, Yves Sirois and Rik Yoshida for the careful reading of the
manuscript. Last but not least, I would like to thank the organizers of the {\sc
Beyond99} conference for a very inspiring meeting.


\renewcommand{\baselinestretch}{0.97}
\bibliographystyle{bib/bsm}
{
\raggedright
\bibliography{bib/bsm99}
\leftline{\small $^a$ XXIX International Conference on High Energy Physics,
                      Vancouver}
\leftline{\small $^b$ International Europhysics Conference on High Energy 
                      Physics, Tampere}
}

\newpage

%
%
\end{document}